\documentclass[aps,preprint,superscriptaddress,groupedaddress]{revtex4}  

\usepackage{graphicx}  
\usepackage{dcolumn}   
\usepackage{bm}        
\usepackage{amssymb}   

\usepackage{xspace}
\usepackage{amsmath}
\usepackage{abbrevs}
\usepackage[]{units}	
\usepackage{color}

\usepackage{float} 
\usepackage{wrapfig} 
\usepackage{multirow} 
\usepackage{rotating} 

\newcommand{\bra}[1]{\langle #1 |}
\newcommand{\ket}[1]{| #1 \rangle}
\newcommand{\braket}[2]{\langle #1 | #2  \rangle}

\begin{document}

\title{Realistic sub-Rayleigh imaging with phase-sensitive measurements}
\author{Kent A.~G. Bonsma-Fisher}
\email[]{kent.bonsma-fisher@nrc.ca}
\affiliation{Centre for Quantum Information \& Quantum Control, Department of Physics, University of Toronto, 60 St. George St, Toronto, Ontario, Canada, M5S 1A7}
\affiliation{National Research Council of Canada, 100 Sussex Drive, Ottawa, Ontario, Canada, K1A 0R6}
\author{Weng-Kian Tham}
\affiliation{Centre for Quantum Information \& Quantum Control, Department of Physics, University of Toronto, 60 St. George St, Toronto, Ontario, Canada, M5S 1A7}
\author{Hugo Ferretti}
\affiliation{Centre for Quantum Information \& Quantum Control, Department of Physics, University of Toronto, 60 St. George St, Toronto, Ontario, Canada, M5S 1A7}
\author{Aephraim M. Steinberg}
\email[]{steinberg@physics.utoronto.ca}
\affiliation{Centre for Quantum Information \& Quantum Control, Department of Physics, University of Toronto, 60 St. George St, Toronto, Ontario, Canada, M5S 1A7}
\affiliation{Canadian Institute For Advanced Research, 180 Dundas St. W., Toronto, Ontario, Canada, M5G 1Z8}

\date{\today}

\begin{abstract}
As the separation between two emitters is decreased below the Rayleigh limit, the information that can be gained about their separation using traditional imaging techniques, photon counting in the image plane, reduces to nil. 
Assuming the sources are of equal intensity, Rayleigh's ``curse'' can be alleviated by making phase-sensitive measurements in the image plane. 
However, with unequal and unknown intensities the curse returns regardless of the measurement, though the ideal scheme would still outperform image plane counting (IPC), i.e., recording intensities on a screen.
We analyze the limits of the SPLICE phase measurement scheme as the intensity imbalance between the emitters grows.
We find that SPLICE still outperforms IPC for moderately disparate intensities. 
For larger intensity imbalances we propose a hybrid of IPC and SPLICE, which we call ``adapted SPLICE'', requiring only simple modifications.
Using Monte Carlo simulation, we identify regions (emitter brightness, separation, intensity imbalance) where it is advantageous to use SPLICE over IPC, and when to switch to the adapted SPLICE measurement.
We find that adapted SPLICE can outperform IPC for large intensity imbalances, e.g., 10,000:1, with the advantage growing with greater disparity between the two intensities.
Finally, we also propose additional phase measurements for estimating the statistical moments of more complex source distributions. 
Our results are promising for implementing phase measurements in sub-Rayleigh imaging tasks such as exoplanet detection. 
\end{abstract}
 
\maketitle

\noindent Rayleigh's criterion places a limit on resolving closely-separated objects~\cite{Rayleigh1879}. 
As the separation between two incoherent point sources becomes smaller, image plane counting (IPC) measurements break down.
More explicitly, the uncertainty in any unbiased estimation of the separation diverges as the separation is decreased below the Rayleigh criterion, which is given by the width of their point spread functions (PSF). 
Recently, strategies have emerged which beat Rayleigh's criterion by using phase information of the two incoherent electromagnetic fields in the image plane, information which is not available when using IPC~\cite{Tsang2016, Lupo2016, Nair2016a, Nair2016b, Rehacek2017, Tsang2018, Lu2018}. 
This phase information allows for the distance between two equal-intensity sources to be estimated with finite variance even as their separation becomes arbitrarily small. 
For a Gaussian PSF, one way to access this phase amounts to making projective measurements in the Hermite-Gauss (HG) position basis.  
A recent flurry of experimental works have used this technique to demonstrate an advantage over IPC. 
In these works the HG projective measurements were performed in various ways including spatial light modulation~\cite{Paur2016}, homodyne detection~\cite{Yang2016}, self-interference~\cite{Tang2016}, or a phase mask followed by single-mode fiber coupling~\cite{Tham2017}. 
Further developments in this rapidly-emerging field have addressed the analogous problem in the spectro-temporal domain~\cite{Donohue2018},  estimating the angular and axial separation of the two point sources~\cite{Napoli2019}, studying the two-~\cite{Ang2017} and three-~\cite{Yu2018} dimensional cases, and by employing Hong-Ou-Mandel interference~\cite{Parniak2018}.
In light of these discoveries, scientists have moved to modernize Rayleigh's criterion~\cite{Tsang2019, Zhou2019}.

Until recently these works have assumed the intensities of the two point sources to be equal. 
It has been shown~\cite{Rehacek2017a, Rehacek2018} that when this assumption is relaxed -- the two sources have unknown, disparate intensities -- that the quantum Cram\'{e}r-Rao bound for an unbiased estimate of the separation diverges as separation decreases below the Rayleigh limit~\footnote{In this work, the Cram\'{e}r-Rao bound refers to the lower bound on variance of an unbiased estimator.}.
However, this divergence happens quadratically slower than the Cram\'{e}r-Rao bound (CRB) for IPC, suggesting that measurement schemes exist which still give an advantage over directly imaging the sources. 
Any further information to be gained must then be in the phase of the incoming light.

In this work we show that the SPLICE measurement scheme, introduced in Ref.~\cite{Tham2017}, can be adapted to estimate the separation of unequal intensity sources. 
We first outline where SPLICE finds an advantage over IPC for estimating the separation of equal-intensity emitters, asking what the lowest resolvable separation is with each technique. 
In the unequal intensity case we find, somewhat surprisingly, that SPLICE can still resolve much lower separations than IPC, within a given error threshold, up until some critical intensity imbalance. 
Beyond that, SPLICE can be adapted with additional projective measurements. We outline these measurements to estimate the variance and skew of the spatial distribution of light from which both the separation and relative intensities can be calculated.
We evaluate the performance of the scheme by looking at the CRB and also Monte Carlo simulations, finding that the adapted SPLICE technique can resolve lower separations than IPC for imbalanced intensities. Moreover, the gap between SPLICE and IPC only grows as the relative intensities between the emitters gets more extreme. 
Following work by Tsang~\cite{Tsang2017}, we also outline a series of SPLICE-inspired measurements to estimate up to the tenth statistical moment for an arbitrary distribution of point sources. 

\section{Super-resolved position localization by inversion of coherence along an edge (SPLICE)}
\label{sec:splice}

\begin{figure}[t!]
\center{\includegraphics[width=0.5\linewidth]{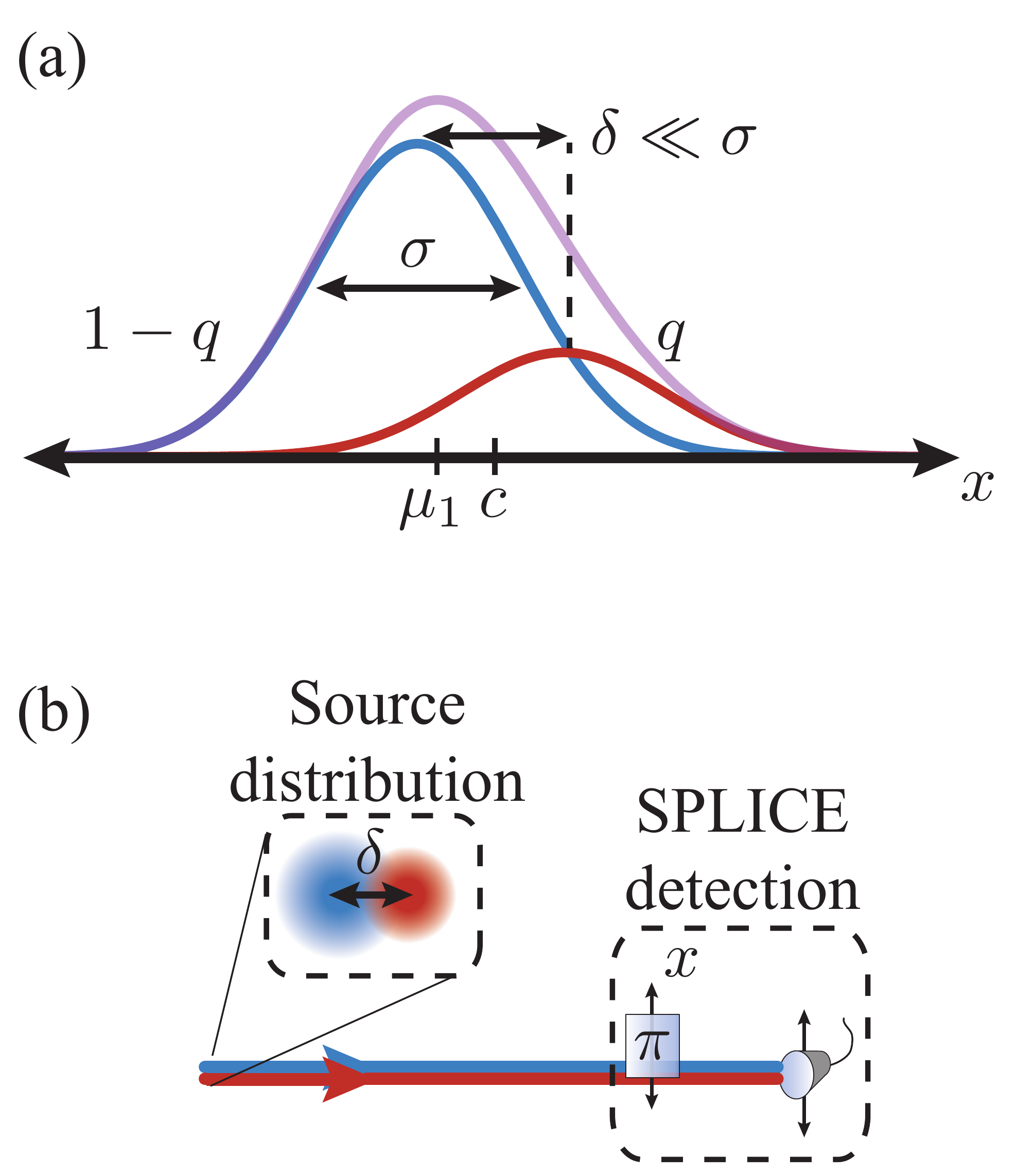}}
\caption{ (a) The intensity distribution from two light sources separated by $\delta$, much smaller than the PSF width $\sigma$. Due to unequal intensities, $q$ and $1-q$, the mean of the distribution, $\mu_1$, is displaced from the geometric centre $c$. (b) The SPLICE detection scheme for measuring the 2nd and 3rd moments along the $x$-axis uses a translatable $\pi$-phase shifter and single-mode fiber. }
\label{fig:twosources}
\end{figure}

\begin{figure}
\center{\includegraphics[width=0.75\linewidth]{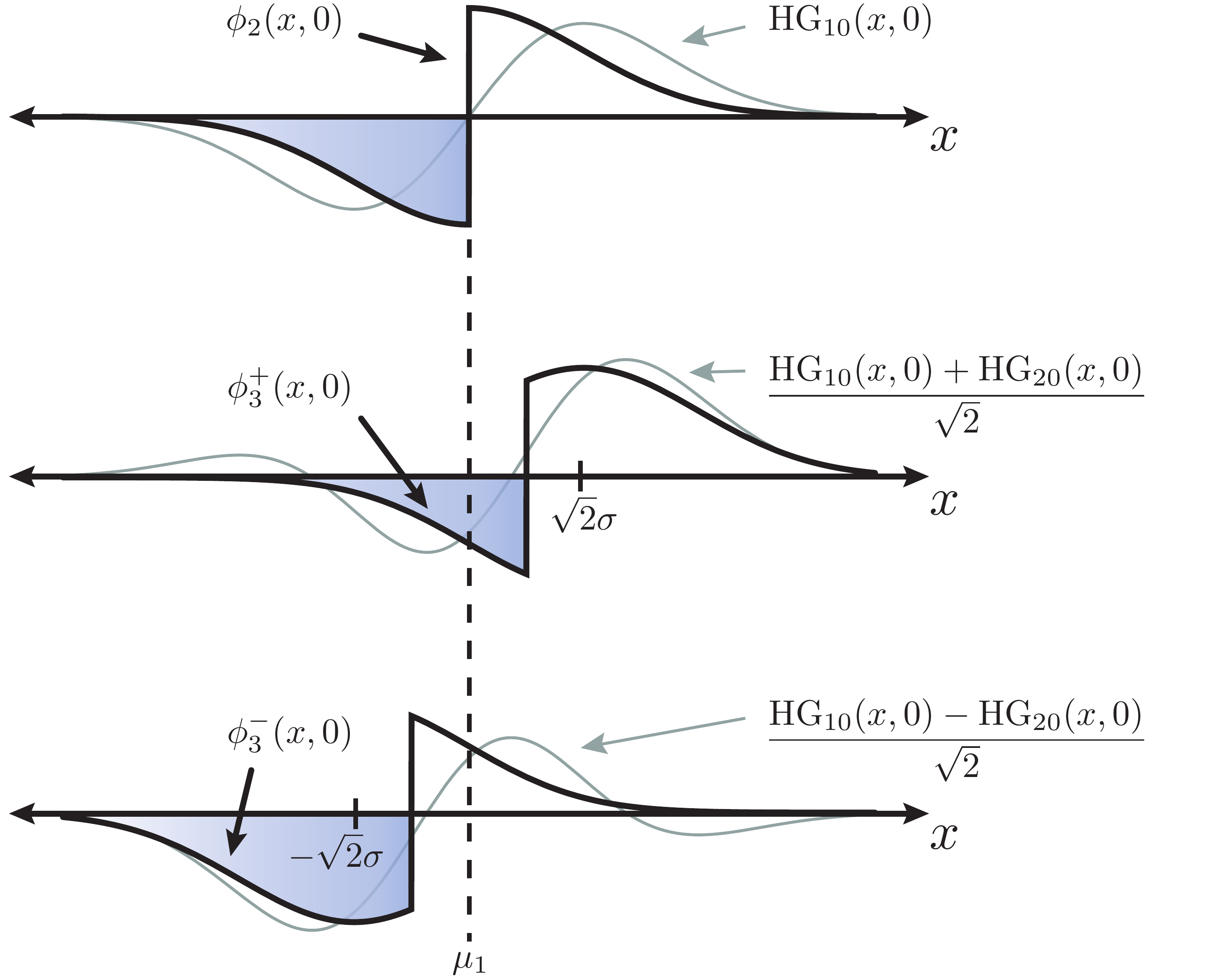}}
\caption{Projections HG$_{10}$, and (HG$_{10} \pm $ HG$_{20}$)$/\sqrt{2}$  (outlined in gray) are approximated by a glass slide, which imparts a $\pi$ phase-shift, followed by a single-mode fiber. To perform adapted SPLICE measurements, the $\pm$ measurements for the skew, the phase-shifter and fiber are translated along the $x$-axis by $\pm \sigma/\sqrt{2}$ and $\pm \sqrt{2} \sigma$, respectively. Since the geometric centre is unknown, measurements are aligned are relative to the mean, $\mu_1$. }
\label{fig:TEM1and2}
\end{figure}

We begin by reviewing the SPLICE measurement scheme introduced by Tham \emph{et al.}~\cite{Tham2017} for the case of equal intensity emitters.
The SPLICE measurement approximates a projection on the HG$_{10}$ mode using a glass edge, which induces a $\pi$ phase-shift on half of the transverse profile of the incoming light, followed by a single-mode fiber (see Fig.~\ref{fig:TEM1and2}).  
We consider two incoherent point sources separated by a distance $\delta$ along the $x$-axis. Each source is transformed by a Gaussian point spread function $| \psi(x,y) |^2$ with width $\sigma$, such that if an emitted photon is equally likely to have come from either source it can be described by the density matrix
\begin{eqnarray}
\rho &=&  \frac{1}{2} \ket{\psi_+}\bra{\psi_+} + \frac{1}{2} \ket{\psi_-}\bra{\psi_-} ,
\label{eq:rho} 
\\
\ket{\psi_\pm} &=& \iint dx dy \hspace{0.1cm} \psi(x \mp \delta/2,y ) \ket{x,y},
\nonumber
\end{eqnarray}
where $\psi(x,y) = \frac{1}{\sqrt{2\pi \sigma^2}} \text{exp} \left \{-\frac{x^2 + y^2}{4\sigma^2} \right \}$. The SPLICE projection $\ket{\phi_2}$ has the form 
\begin{eqnarray}
\label{eq:oldsplicedef}
\ket{\phi_2} &=& \frac{1}{\sqrt{2 \pi \sigma^2}} \iint dx dy \hspace{0.1cm} e^{-\frac{x^2 + y^2}{4\sigma^2}} \text{sgn} ( x ) \ket{x,y}.
\end{eqnarray}
The probability of a successful photon detection is $p_2 = \bra{\phi_2}\rho\ket{\phi_2}  \approx \frac{\delta^2}{8 \pi \sigma^2} + \mathcal{O}(\frac{\delta^4}{\sigma^4})$ for $\delta \ll \sigma$. 
The separation is then estimated by $\hat \delta_\text{SPL} = \sqrt{8\pi \sigma^2 k/N}$, where $k$ is the number of photons detected in a given time interval with an average incident photon flux $N$, i.e., $\left \langle k \right \rangle = N p_2$. With IPC, one records the transverse positions of $N$ incident photons on a screen, which are drawn from the intensity distribution $ \mathcal{I} = \frac{1}{2} | \braket{x}{\psi_+} |^2 + \frac{1}{2} | \braket{x}{\psi_-} |^2$. 
The separation can be estimated by subtracting $\sigma^2$ from the estimate of the variance, $\hat \delta_\text{IPC} = 2\sqrt{( \hat v - \sigma^2)}$.

\begin{figure*}
\center{\includegraphics[width=0.8\linewidth]{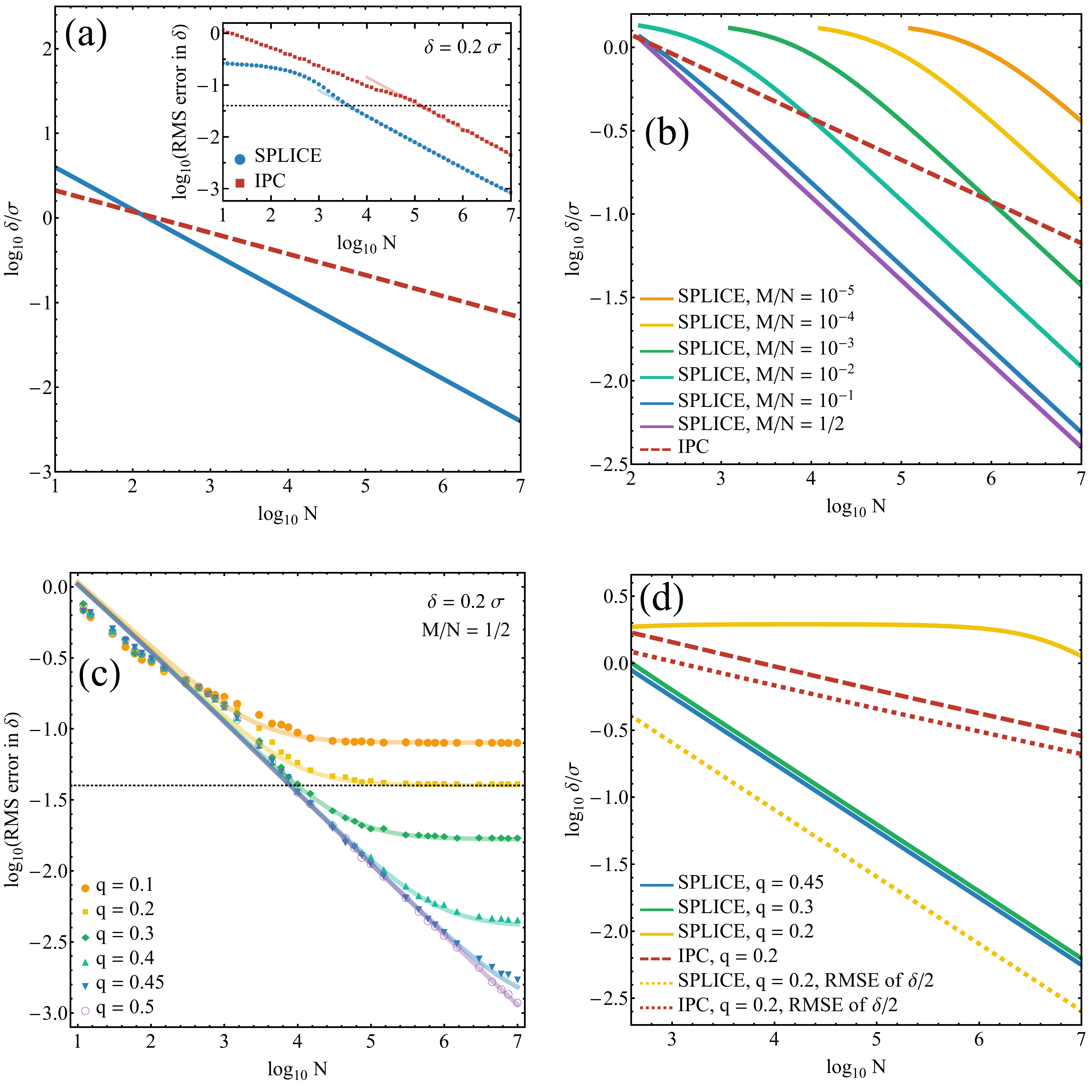}}
\caption{ (a) Lowest $\delta$ that can be distinguished from 0 by 5 standard deviations (RMS error $\leq \delta / 5$) using unadapted SPLICE (solid blue), or IPC (dashed red). 
Lowest resolvable $\delta$ scales as $N^{-1/2}$ for SPLICE but as $N^{-1/4}$ for IPC. 
Inset: Monte Carlo simulations showing the RMS error in $\hat \delta$ when $\delta = 0.2 \sigma$.
Semi-transparent lines show the Cram{\'e}r-Rao bounds for each method.
Dotted black line marks $\delta/5$. 
(b) With equal intensities, but an unknown mean, first detect $M$ photons with IPC to estimate the mean and align the SPLICE apparatus. 
Shown is the lowest $\delta$ estimated with RMS error $\leq \delta /5$ for various fractions $M/N$.
(c) RMS error in $\delta$ for SPLICE when emitters have unequal intensities ($q$ and $1-q$). Here, $\delta = 0.2\sigma$ and $M = N/2$. Dotted black line marks an RMS error of $\delta/5$.
(d) Lowest separation estimated with SPLICE and IPC for unequal intensity emitters. For $q<0.2$ SPLICE cannot achieve an RMS error of $\delta/5$ for any $\delta$. For $0.2<q<0.5$, or a larger RMS error threshold (dotted lines), the lowest resolvable $\delta$ scales as $N^{-1/2}$ for SPLICE and $N^{-1/4}$ for IPC.
}
\label{fig:oldsplice}
\end{figure*}

Our primary concern is to identify the lowest value of $\delta$ that, using either SPLICE or IPC, can be estimated to within some  root-mean-square (RMS) error threshold.
Throughout this work we choose an error threshold of $\delta/5$.
This is done by comparing the Cram{\'e}r-Rao bounds for the two measurement techniques. 
The CRB, the mean-squared error for an unbiased estimator, for a projective phase measurement such as SPLICE is independent of $\delta$ in the sub-diffraction regime.
We stress that this is only true for the equal-intensity case. 
For SPLICE the CRB is $\frac{2\pi \sigma^2}{N}$, whereas for IPC the CRB is $\frac{8\sigma^4}{N \delta^2}$.
These are plotted in the inset of Fig.~\ref{fig:oldsplice}(a) along with the RMS error of $\hat \delta_\text{SPL}$ and $\hat \delta_\text{IPC}$ from Monte Carlo simulations (see Appendix~\ref{sec:MC} for details). 
For low $N$, neither the SPLICE nor IPC estimator agrees with its respective CRB due to estimator bias. 
In the biased regime, the standard deviation of the SPLICE estimate is constant.  
This biased regime extends through $0 < \delta^2 \lessapprox  8 \pi \sigma^2 / N$, and becomes unbiased approximately when the expected number of detected photons is 1. This is discussed further in Appendix~\ref{sec:numerical}.
The standard deviation of the IPC estimate scales as $N^{-1/4}$ in the biased regime. The IPC biased regime extends through $0< \delta^4 \lessapprox 64\sigma^4/N$, and becomes unbiased approximately when the error in the $\mathcal{I}$ variance estimate becomes smaller than $(\delta/2)^2$. 

For low error tolerances (black dotted line in Fig.~\ref{fig:oldsplice}(a) inset) it suffices to simplify our discussion to only the CRB scaling. 
That is to say, while in the biased regimes neither estimator will reach the desired error threshold. 
In Fig.~\ref{fig:oldsplice}(a) we plot the line where the square-root of the CRB is equal to an error tolerance of $\delta/5$. 
This represents the lowest resolvable separation for each measurement technique at a given source brightness $N$. 
For separations $\delta < \sigma$ we see that SPLICE can resolve smaller separations than IPC. 
Importantly, the smallest separation resolved by SPLICE scales as $N^{-1/2}$ compared to a $N^{-1/4}$ scaling for IPC. 

In a more realistic setting, the geometric center of the source distribution is not known a priori, and it is unclear of where to position the edge of the SPLICE phase-shifter. 
Ref.~\cite{Tsang2016} analysed the setting where a binary SPADE apparatus was aligned based on an initial IPC estimate. 
Here, we do the same for our (unadapted) SPLICE apparatus. 
As we will discuss in the next section, unequal intensities introduces the additional complication that even an IPC estimate of the geometric centre is biased estimate from the true value. 
A goal of the present work is to identify how SPLICE compares to IPC in this scenario, and whether an adapted SPLICE measurement can yield an advantage over IPC.
Before discussing this further, it is helpful to revisit the equal-intensity scenario where the geometric centre is not known a priori. 
The strategy we consider is to estimate the mean position ($\bar x$) of the intensity distribution using IPC, to which we align the SPLICE apparatus.
A misalignment from the mean $\Delta\mu = \bar x - \mu_1$ affects the probability of a detection using SPLICE as $p_2 \approx \frac{\delta^2}{8 \pi \sigma^2} + \frac{\Delta \mu^2}{2 \pi \sigma^2} + \mathcal{O} \left (\frac{\delta^4}{\sigma^4} \right)$.
The misalignment $\Delta \mu$ is a normally distributed variable centred at 0 with variance $\sigma^2 / M$, where $M$ is the number of photons used for mean estimation. 
We evaluate the estimator $\hat \delta_\text{SPL}$ to compare with IPC, and in Fig.~\ref{fig:oldsplice}(b) we plot the smallest separation that can be estimated to within $\pm \delta/5$ for both IPC and SPLICE.
This is plotted for different fractions of the total photon input used for the mean estimation, $M/N$.
The $N^{-1/2}$ scaling for resolving the lowest separation persists using SPLICE but, due to poor estimation of the mean position at low photon numbers, the scaling does not start until large $N$.
With larger fractions of photons allocated for mean estimation, the bias in the SPLICE measurement caused by misalignment is mitigated and the $N^{-1/2}$ scaling begins at lower values of $N$. 

\section{Unadapted SPLICE with unequal-intensity emitters}
\label{sec:unequal}

Relaxing the equal-intensity assumption, we have the scenario shown in Fig.~\ref{fig:twosources}(a). The two emitters are separated by $\delta$ about a centre point $c$. 
A given photon with probability $q$ of coming from the right emitter, and $1-q$ from the left, is described by the density operator
\begin{eqnarray}
\rho &=&  q \ket{\psi_+}\bra{\psi_+} + (1-q) \ket{\psi_-}\bra{\psi_-}.
\label{eq:rho2} 
\end{eqnarray}
There are now three unknown parameters, $\{ \delta, q, c \}$. 
It should be stated here that if $c$ is known, one can align the SPLICE measurement about $c$ and recover the same advantage as in the equal-intensity case. 
However, estimating $c$ is not trivial since, when $q \neq 1/2$, $c$ is offset from the mean of the distribution by an amount related to the skew. 
The most na\"{i}ve strategy is to align the SPLICE measurement about the mean, $\mu_1 = c - (1-2q)\frac{\delta}{2}$, which is best estimated using IPC. Due to symmetry, for the rest of this work we take $0 \leq q \leq 1/2$.

With the SPLICE projection the same as in the previous section, the probability of a photon detection is now 
\begin{eqnarray}
    p_2 &=& \bra{\phi_2}\rho\ket{\phi_2} 
    \nonumber \\
    &=& 
    \frac{q(1-q) \delta^2}{2 \pi \sigma^2} 
    + \frac{2q(1-q)(1-2q) \delta^3}{3\pi \sigma^4} \Delta \mu
    + \left (  1 - 2q(1-q)\frac{\delta^2}{\sigma^2}  \right ) \frac{\Delta \mu^2}{2 \pi \sigma^2} 
    + \mathcal{O} \left(\frac{\delta^4}{\sigma^4} \right ).
    \label{eq:qprob}
\end{eqnarray}
The SPLICE estimator, $\hat \delta_\text{SPL} = \sqrt{8 \pi \sigma^2 k / N}$, now includes more terms depending on $q$. When the misalignment $\Delta \mu$ can be neglected, for example when $N$ is large, the estimator returns $\hat \delta_\text{SPL} = 2\delta \sqrt{q(1-q)}$. 
Fig.~\ref{fig:oldsplice}(c) shows the RMS error for $\hat \delta_\text{SPL}$ as a function of $N$ when $\delta = 0.2\sigma$ for various values of $q$.
We plot the results of Monte Carlo simulations as well a numerically solved analytical expression (see Appendices~\ref{sec:MC} and \ref{sec:numerical} for details). 
Note that for these simulations we take half of the incident light to estimate the mean ($M/N = 1/2$).
We see that as the emitters deviate from equal intensity, $q=1/2$, the estimate results in a constant error at high $N$. This error comes from the fact that the SPLICE estimator explicitly depends on $q$, resulting in a bias of $\delta(1-2\sqrt{q(1-q)})$ at large $N$. 
For a given RMS error threshold $\epsilon$, there is a critical intensity value above which SPLICE can resolve separations with error $\leq \epsilon$. The critical intensity is $| q_c - \frac{1}{2}| = \frac{1}{2}\sqrt{\frac{\epsilon}{\delta}(2-\frac{\epsilon}{\delta})}$.
For our chosen error threshold $\epsilon = \delta/5$ the critical value of $q_c$ is 1/5.
This is outlined in Fig.~\ref{fig:oldsplice}(d), where we again plot the lowest resolvable separations for an error threshold of $\delta /5$.
In the regime where SPLICE still functions, $1/5 < q < 1/2$, it outperforms IPC: SPLICE has a $N^{-1/2}$ scaling compared to a $N^{-1/4}$ scaling for IPC. 
By allowing for bias in the SPLICE estimator it is still possible to reduce the RMS error below the threshold.
Given that the SPLICE measurement was constructed with the assumption that the emitters are equally bright, is it somewhat surprising that SPLICE can still function at moderately disparate intensities.
In the next section we discuss how the SPLICE measurement can be adapted for more extreme values of $q$.

\section{Adapting SPLICE for unequal-intensity emitters}
\label{sec:newsplice}

\noindent In the equal-intensity scenario the SPLICE measurement amounts to estimating the variance (2nd central moment) of the spatial distribution of the incident light, from which the separation can be calculated.  
After relaxing the equal-intensity assumption the variance no longer gives full information and, as one might expect from Fig.~\ref{fig:twosources}(a), the skew (3rd central moment) of the distribution becomes relevant. 
By translating the phase shifter and fiber transversely, with respect to the direction of light propagation, we can perform two additional measurements to estimate the skew.
With estimates of the variance ($\mu_2 = q(1-q)\delta^2$) and skew ($\mu_3 = q(1-q)(1-2q)\delta^3$), both the separation ($\delta$) and relative intensities ($q$) can be calculated. 
We introduce the projective measurements $\ket{\phi^\pm_3}$ to estimate $\mu_3$. Higher moments of the source distribution can also be measured with additional phase shifters across the transverse direction (see Section~\ref{sec:higher}).
The measurements for $\mu_3$ are described by the projectors
\begin{eqnarray}
\label{eq:splicedef}
\ket{\phi_3^\pm} &=& \frac{1}{\sqrt{2 \pi \sigma^2}} \iint dx dy \hspace{0.1cm} \phi_3^\pm(x,y) \ket{x,y},
\nonumber \\
\phi^\pm_3(x,y) &=& e^{-\frac{(x \mp \sqrt{2}\sigma)^2 + y^2}{4\sigma^2}} \text{sgn} ( x \mp \sigma/\sqrt{2} ),
\label{eq:phi3}
\end{eqnarray}
which are, like $\ket{\phi_2}$, realized by a $\pi$ phase-shifter and single-mode fiber coupler, each translatable along the $x$-axis. 
The $\ket{\phi^\pm_3}$ projectors are constructed to approximate the iTEM projectors of Ref.~\cite{Tsang2017}, $\left ( \ket{\text{HG}_{10}} \pm \ket{\text{HG}_{20}} \right )/\sqrt{2}$  (see Fig.~\ref{fig:TEM1and2}). 
Expressing $\ket{\phi^\pm_3}$ in the Hermite-Gauss basis we have that 
\begin{eqnarray}
\ket{\phi^\pm_3} &=& \sqrt{\frac{2}{\pi \sqrt{e}}} \left ( \ket{\text{HG}_{10}} \pm \ket{\text{HG}_{20}} \right )  + \frac{1}{2\sqrt{3\pi \sqrt{e}}} \ket{\text{HG}_{30}}  \mp  ...,
\\
\ket{\text{HG}_{mn}} &=& \iint dx dy \hspace{0.1cm} \text{HG}_{mn} (x,y) \ket{x,y},
\nonumber \\
\text{HG}_{mn} (x,y) &=& \frac{\text {He}_m \left (\frac{x}{\sigma} \right ) \text{He}_n  \left (\frac{y}{\sigma} \right ) }{\sqrt{2\pi \sigma^2 m! n!}}  \text{exp} \left \{- \frac{x^2+y^2}{4\sigma^2} \right \},
\label{eq:temdef}
\end{eqnarray}
where $\text{He}_n (x)$ is the $n^\text{th}$ probabilists' Hermite polynomial. 
For a Gaussian PSF the measurement success probability $\bra{\text{HG}_{m0}} \rho \ket{\text{HG}_{n0}}$ is proportional to the $n+m^\text{th}$ statistical moment, $\mu_{m+n}$, to leading order in $\delta/\sigma$. 
For the 2nd and 3rd moments we have that $\mu_2 \propto \bra{\phi_2}\rho \ket{\phi_2}$, and $\mu_3 \propto \bra{\phi^+_3}\rho \ket{\phi^+_3} - \bra{\phi^-_3}\rho \ket{\phi^-_3}$. In the latter case, the $\mu_2$ and $\mu_4$ terms are canceled in the subtraction. 
An interesting fact which emerges here is that the SPLICE modes chosen in Eq.~\ref{eq:phi3} do not give maximal overlap with $\left ( \ket{\text{HG}_{10}} \pm \ket{\text{HG}_{20}} \right )/\sqrt{2}$. 
That is to say, there is another position for the phase-shifter and fiber coupler which achieves higher overlap. 
However, making this choice also results in non-zero overlap with $\ket{\text{HG}_{00}}$. 
This in turn gives a contribution from $\mu_1$ which does not cancel in the subtraction and dominates over the $\mu_3$ term.  
In constructing the SPLICE measurements it is crucial to ensure zero overlap with all lower HG modes.

\begin{figure}
\center{\includegraphics[width=0.5\linewidth]{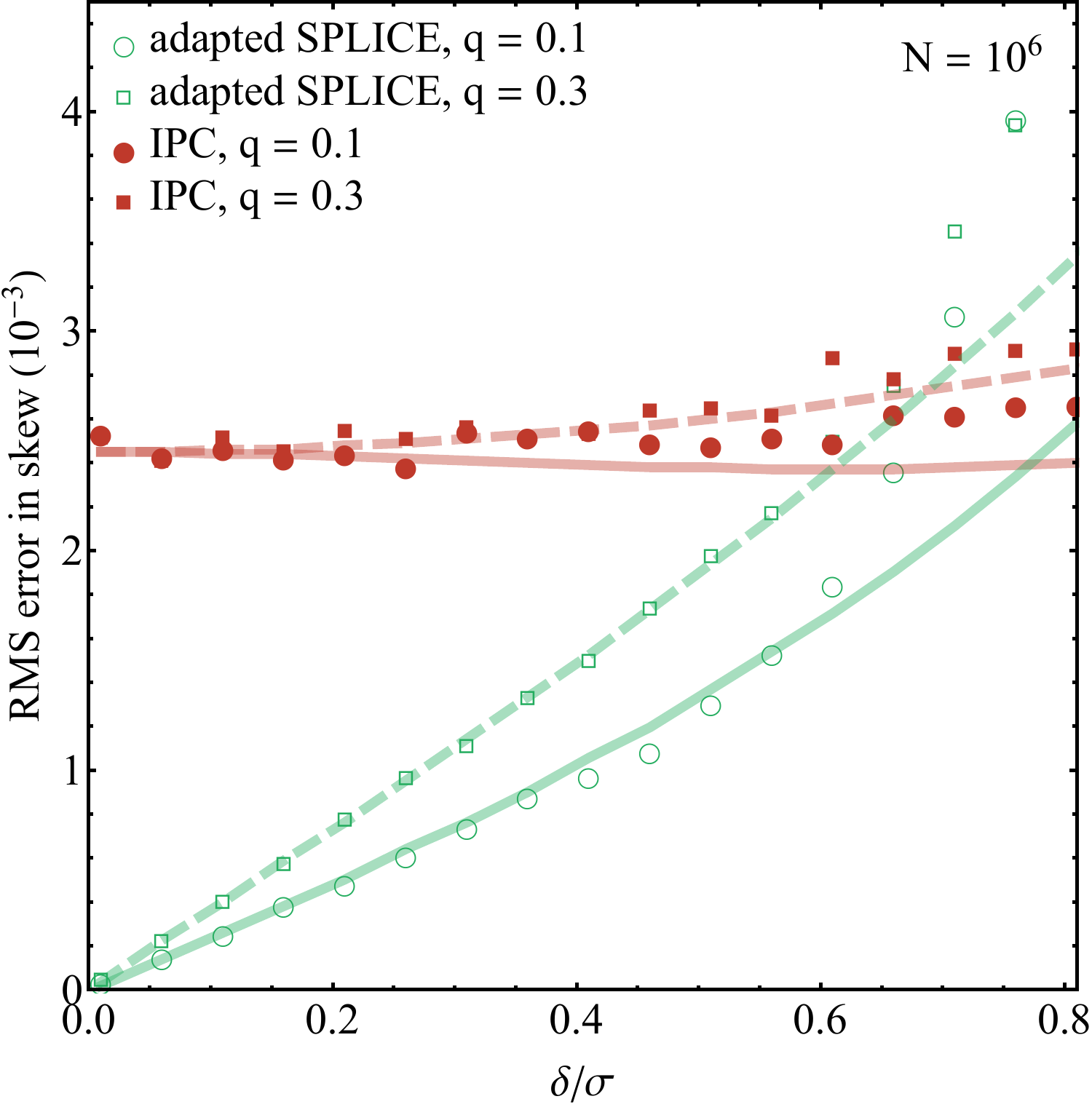}}
\caption{Monte Carlo results for $N=10^6$ input photons and $1,000$ Monte Carlo repetitions.  
Root-mean-squared (RMS) error of skew estimates for $q = 0.1$ (circles) and $q=0.3$ (squares) as a function of separation. 
In the adapted SPLICE scheme (green, hollow) $N/3$ photons are used in the skew measurement, split evenly between the two projections. 
For IPC (red, solid) all $N$ photons can be used for the skew estimate. 
Lines show the CRB for $q = 0.1$ (solid) and $q=0.3$ (dashed) cases. 
As $\delta/\sigma$ decreases, the RMS error in skew given by adapted SPLICE dips well below the constant error given by IPC.
}
\label{fig:rmsskew}
\end{figure}

When the $\ket {\phi^\pm_3}$ projectors are aligned with respect to $\bar x$, misaligned from the mean by $\Delta \mu = \bar x - \mu_1$, the probabilities of photon detection are
\begin{eqnarray}
p^\pm_3 &=& \bra{\phi^\pm_3}\rho \ket{\phi^\pm_3}
\nonumber \\
&=& 
\frac{q(1-q)\delta^2}{2\sqrt{e} \pi \sigma^2}
\pm \frac{q(1-q)(1-2q) \delta^3}{2\sqrt{2e}\pi \sigma^3}
\mp  \frac{3q(1-q) \delta^2}{2\sqrt{2e} \pi \sigma^3} \Delta \mu 
\nonumber \\
&& \quad \quad + \left (
1 - \frac{q(1-q)\delta^2}{2\sigma^2}
\right ) \frac{\Delta \mu^2}{2\sqrt{e} \pi \sigma^2}
+ 
\mathcal{O}  \left ( \frac{\delta^4}{\sigma^4} \right ).
\end{eqnarray}
Ideally, if $\Delta \mu \approx 0$, the difference between the two measurements gives $p^+_3 - p^-_3 = \frac{q(1-q)(1-2q)}{\sqrt{2e}\pi} ( \delta / \sigma  )^3 + \mathcal{O}  \left ( \delta^5 / \sigma^5 \right )$.
We then construct the estimators for the separation and intensities from the variance and skew measurements as $\hat v = 2\pi \sigma^2 p_2$, and $\hat s = \sqrt{2e}\pi \sigma^3 ( p^+_3 - p^-_3)$, respectively, which can be inverted to find estimates for $\delta$ and $q$, 
\begin{eqnarray}
\hat \delta &=&  \sqrt{ 4 \hat v  + (\hat s/ \hat v)^2},
\label{eq:deltaest}
\\ \nonumber \\
\hat q &=& \frac{1}{2} \left ( 1 -  \frac{\hat s/ \hat v }{\sqrt{ 4 \hat v  + (\hat s/ \hat v)^2}} \right ) .
\label{eq:qest}
\end{eqnarray}
Given $k$ photons detected during the $\ket{\phi_2}$ measurement, and $j^\pm$ detected during the $\ket{\phi^\pm_3}$ measurements, the full adapted SPLICE separation estimator is 
\begin{equation}
    \hat \delta_\text{aSPL} = \sqrt{ \frac{8\pi \sigma^2 k}{N_2} + \left [ \frac{\sqrt{2e}\pi \sigma^3 (j^+/N_3^+ - j^-/N_3^-)}{2\pi \sigma^2 k/N_2} \right ]^2 },
\end{equation}
where $N_2, N_3^\pm$ incident photons are taken for the respective measurements. 
For the remainder of this work, we allocate incident photons evenly between the mean, variance, and skew measurements such that $N/3 = M = N_2 = 2N_3^+ = 2N_3^-$.

We perform Monte Carlo simulations (see Appendix~\ref{sec:MC}) to evaluate this adapted SPLICE measurement for estimating $\delta$ and $q$. 
The results are compared against simulations of IPC, as well its CRB.
We will first compare estimations of the skew using the adapted SPLICE measurement. 
Figure~\ref{fig:rmsskew} shows the RMS error in skew estimates for cases where $q=0.1$ and $q=0.3$. 
We see that as $\delta/\sigma$ gets smaller, the error in the skew from IPC becomes constant at $\sqrt{3! / N}$. 
Meanwhile, the adapted SPLICE estimate shows an RMS error that decreases with the separation.
This was first discussed in Ref.~\cite{Tsang2017}, where it was shown that projective phase measurements for the statistical moments of a source distribution, in the style of SPADE~\cite{Tsang2016} or SPLICE, provide unbiased estimates and have substantially lower Cram{\' e}r-Rao bounds than IPC in the sub-Rayleigh region. 

\begin{figure}
\center{\includegraphics[width=\linewidth]{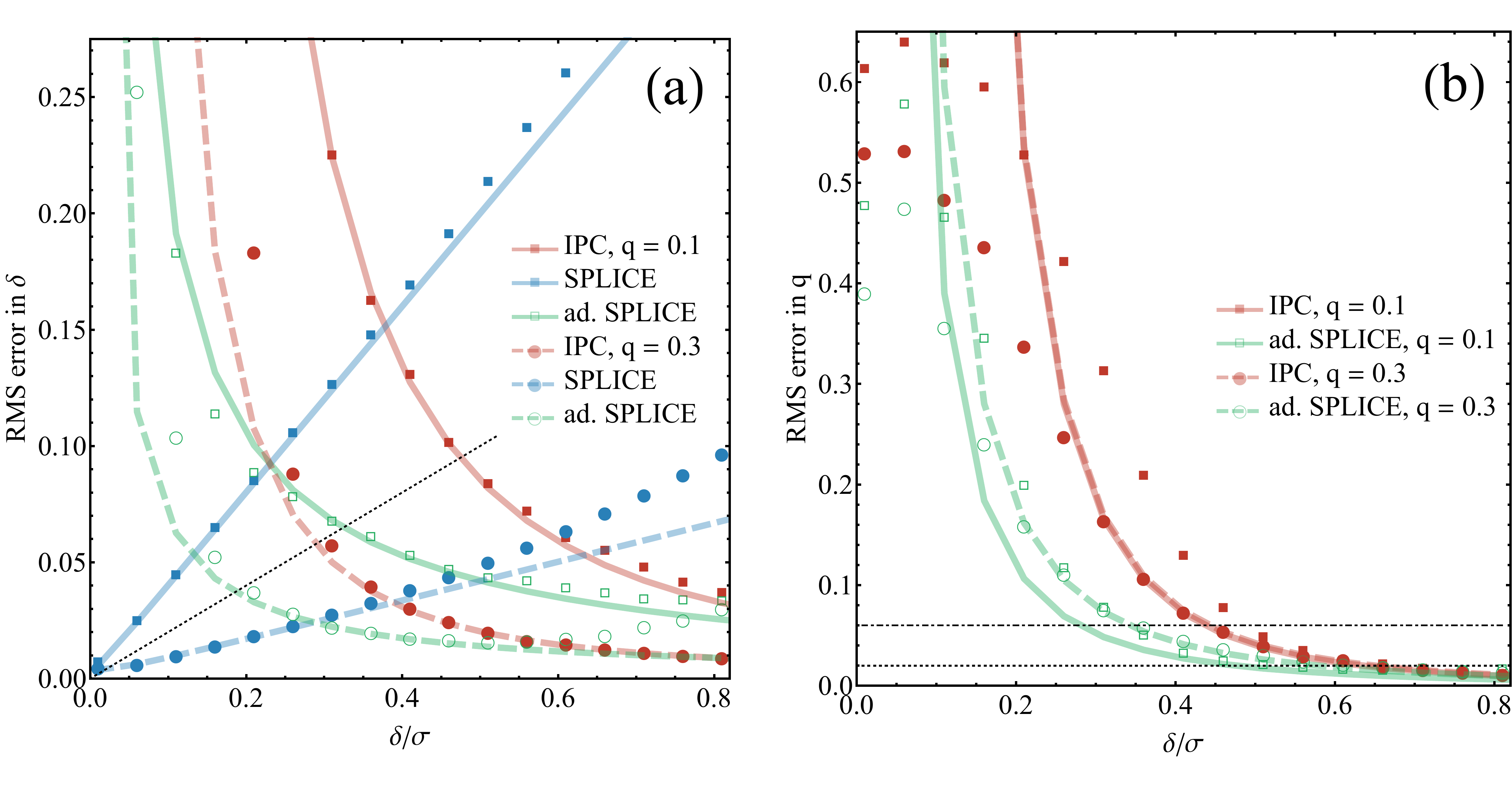}}
\caption{Monte Carlo results for $N=10^6$ photons and $1,000$ Monte Carlo repetitions.
RMS error of (a) $\delta$, and (b) $q$ estimates as a function of separation when $q=0.1$ (squares) and $q=0.3$ (circles) for IPC (solid red), SPLICE (solid blue) and adapted SPLICE (hollow green). 
Solid ($q=0.1$) and dashed ($q=0.3$) lines show the CRB for IPC and adapted SPLICE. 
(a) For SPLICE, solid and dashed lines are numerically calculated RMS errors (see Appendix~\ref{sec:numerical}), linear in $\delta$ for $N \gg 2\pi\sigma^2 / \delta^2 q(1-q)$.
Dotted black line marks an RMS error of $\delta/5$. 
While the error for adapted SPLICE diverges (whereas it does not for SPLICE), there are values of $\delta$ (for $q=0.1$) where SPLICE cannot achieve an RMS error below $\delta/5$, but where adapted SPLICE can. 
(b) Since SPLICE does not provide an estimate of $q$, we only compare IPC and adapted SPLICE.
Monte Carlo results that fall below the CRB lines imply biased measurements. 
Adapted SPLICE can achieve RMS errors below thresholds of $q/5$ (dotted and dot-dashed black lines) for smaller separations than IPC can.  
}
\label{fig:RMSdq}
\end{figure}

Figure~\ref{fig:RMSdq} shows the resulting RMS errors when estimating $\delta$, and $q$, by either IPC, SPLICE, or adapted SPLICE for the same cases of $q$ as above.
Here, we plot the results of Monte Carlo simulations for the three methods, overlaid with the relevant CRBs (except in the case of SPLICE (blue) where we instead overlay the numerically found RMS error).
First, we see in Fig.~\ref{fig:RMSdq}(a) that the error when estimating $\delta$ diverges at low separations when using either IPC or adapted SPLICE. 
This was to be expected since it is now known that the quantum CRB for estimating $\delta$ between unequal intensity emitters diverges at low $\delta$~\cite{Rehacek2017a,Rehacek2018}.
However, we also see that the unadapted SPLICE measurement, which assumes equal intensity emitters, does surprisingly well in comparison.
The RMS error does not diverge at low separation.
As seen in the previous section, for values of $q$ above some critical value, i.e., $q_c = 1/5$, one can achieve RMS errors lower than a given error threshold down to very low separations. 
This is surprising given that the quantum CRB diverges at low separations. 
However, the SPLICE estimator, $\hat \delta_\text{SPL}(k) = \sqrt{8 \pi \sigma^2 k /N}$, differs from the estimator in Eq.~\ref{eq:deltaest} primarily by the factor of $s^2/v^2$, which manifests as a bias. 
Then for low skews, which occur for $q \approx 1/2$ or as $\delta \to 0$, the unadapted SPLICE estimator can still achieve a low bias while retaining excellent scaling with $N$.
Once $q < q_c$, the error threshold cannot be reached for any separation even though the RMS error decreases linearly with $\delta$. 
Therefore, for scenarios with large skew, $q < q_c$, we turn to the adapted SPLICE measurement. 
In Fig.~\ref{fig:RMSdq}(a), we see that the RMS error diverges at low separations.
This can be explained by that when adapting the measurement to estimate both $\delta$ and $q$, there are now different physical scenarios that can lead to a low skew: 
two near-equal intensity sources close together ($q\approx 1/2$, $\delta \to 0$), or very disparate intensity sources far apart ($q \to 0$, $\delta \gg 0$). 
This was avoided in the unadapted SPLICE measurement by always assuming that $q = 1/2$. 
Though the RMS error for $\hat \delta_\text{aSPL}$ diverges for low $\delta$, there is still a range of $\delta$ which can be estimated below the error threshold even at low $q$.
Additionally, the adapted SPLICE measurements allow one to make an estimate of $q$ itself, which we show in Fig.~\ref{fig:RMSdq}(b).
We see that the error in $\hat q$ also diverges at low $\delta$.
However, the adapted SPLICE measurement can make estimates of $q$ below a chosen error threshold, e.g., $q/5$, for lower separations than can be achieved by using IPC alone. 

\begin{figure}
\center{\includegraphics[width=0.85\linewidth]{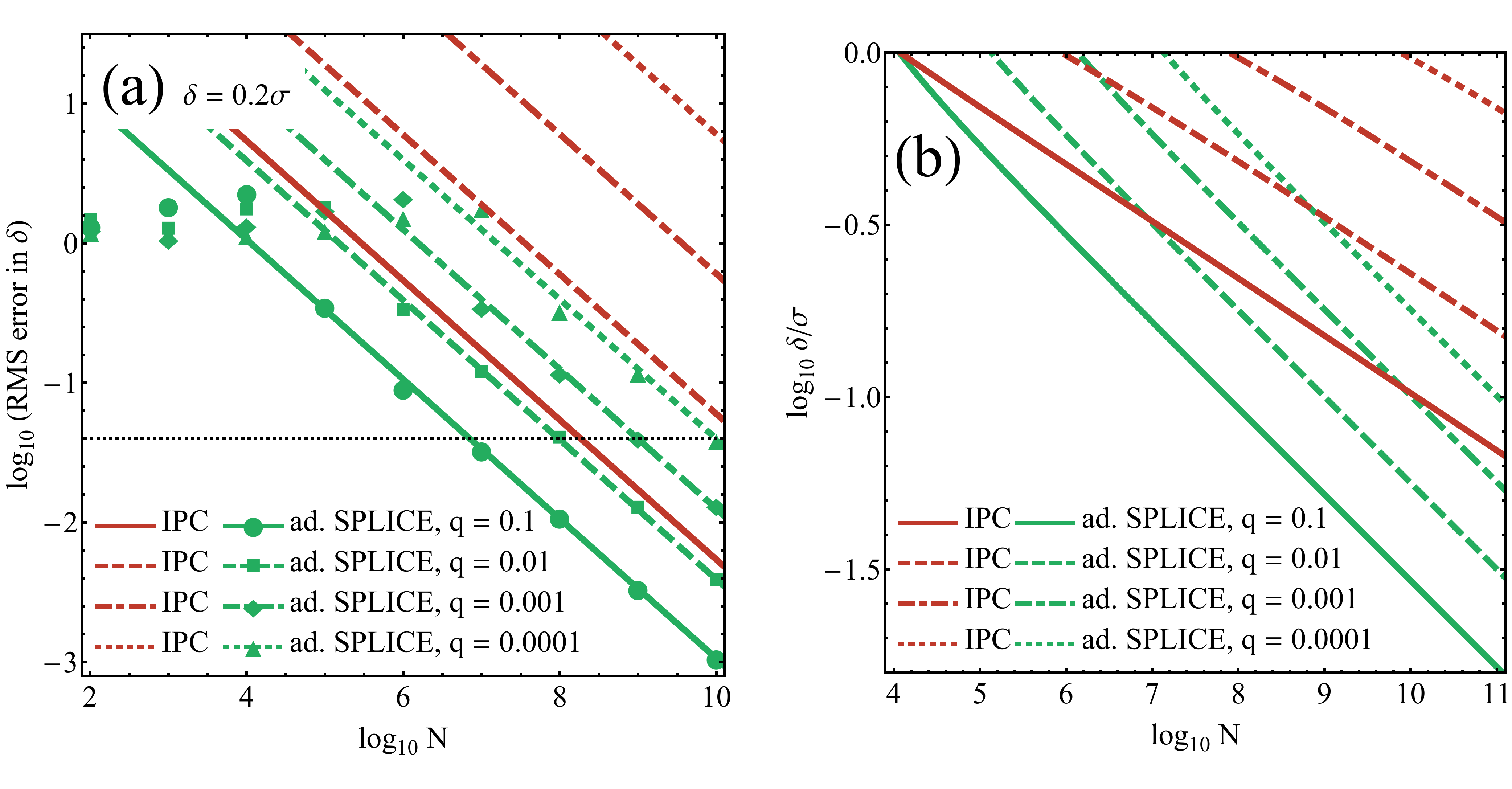}}
    \caption{(a) The RMS error in estimating $\delta$, using the adapted SPLICE (green) or IPC (red), when the actual separation is $\delta = 0.2\sigma$ and one of the emitters is much brighter than the other (fraction given by $q$).  Points show results from Monte Carlo simulations of the adapted SPLICE, allocating photon numbers equally between the mean (IPC), variance and skew measurements. Lines mark Cram{\'e}r-Rao bound for SPLICE and IPC for each value of $q$. The black dotted line shows where the RMS error is equal to $\delta/5$. 
    (b) Separations that can be estimated to within an RMS error of $\delta/5$ with $N$ photons using adapted SPLICE (green) or IPC (red). Lines are calculated from the Cram{\'e}r-Rao bounds in (a). }
\label{fig:extremeq}
\end{figure}

Finally for this section, we investigate a scenario where one emitter is much brighter than the other.
Figure~\ref{fig:extremeq} compares the adapted SPLICE scheme to IPC for estimating $\delta$ when $q$ ranges from $10^{-1}$ to $10^{-4}$. 
In Fig.~\ref{fig:extremeq}(a), Monte Carlo simulations of adapted SPLICE (when $\delta = 0.2\sigma$) show relatively constant RMS errors until $N \sim 2 \pi \sigma^2 / q(1-q) \delta^2$, after which we see behaviour close to the CRB, with the RMS error scaling like $1/\sqrt{N}$. 
Using adapted SPLICE, if $q$ is reduced by an order of magnitude, achieving a fixed RMS error threshold requires an order of magnitude more photons.
Looking at the CRB for IPC in Fig.~\ref{fig:extremeq}(a), a drop in $q$ by an order of magnitude requires an additional two orders of magnitude of photons to achieve the same RMS error threshold. 
This is due to the fact that the IPC estimator is biased until $N \sim \frac{2 \sigma^4}{\delta^4 q^2(1-q)^2}$, below which the standard deviation only scales like $N^{-1/4}$.
Monte Carlo simulations for IPC were computationally intensive and were not performed for input photon numbers of $N > 10^7$, though we found good agreement between simulations and the CRB for $q = 0.1$ when $N = 10^6$ (see Fig.~\ref{fig:RMSdq}(a)).

We compare the CRB of IPC to that of adapted SPLICE in Fig.~\ref{fig:extremeq}(b), where we once again plot the smallest resolvable separation below an RMS error threshold of $\delta /5$.
Note that since the biases of both the adapted SPLICE and IPC estimators are large for few photons, the error threshold will only be reached when the bias is low and  RMS error is dominated by the CRB. 
Crucially, we see that the slopes of these lines are quite different: the lowest resolvable $\delta$ using adapted SPLICE scales approximately as $N^{-1/4}$, whereas using IPC it scales approximately as $N^{-1/6}$.
Furthermore, we see again that the required $N$ for a given $\delta$ scales linearly with $q$ for adapted SPLICE, but quadratically with $q$ for IPC. 
This is primarily determined by the amount of photons required for each scheme to overcome their low-$N$ biases, which are given above. 
The adapted SPLICE scheme shows promising scaling advantages in photon number over IPC for emitters of very different brightness and separations below the Rayleigh limit.

\section{Estimating higher moments}
\label{sec:higher}

\begin{figure*}
\center{\includegraphics[width=0.85\linewidth]{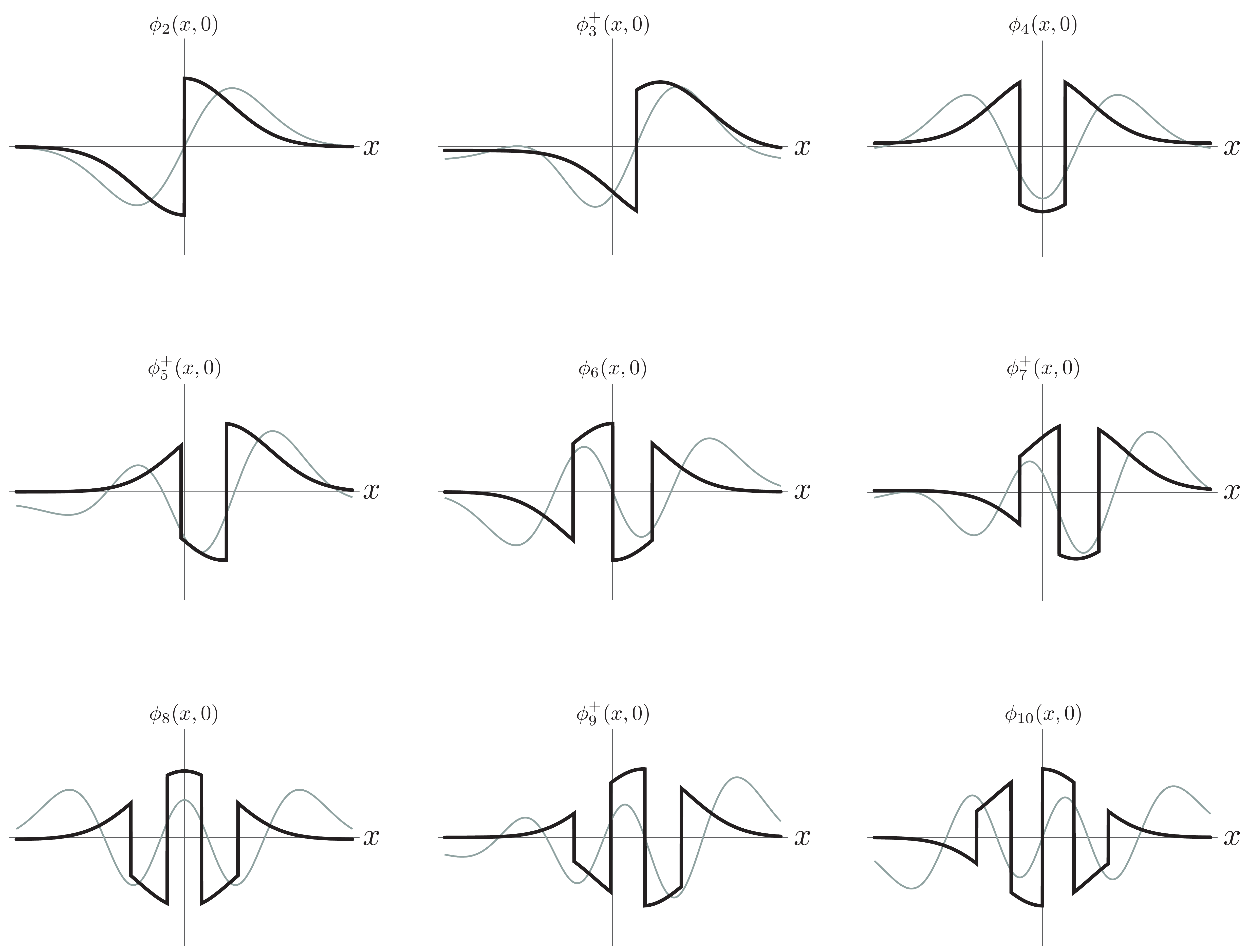}}
\caption{Generalized SPLICE functions $\phi_j(x,0)$ (solid black, Eq.~\ref{eq:splicedef}) are used to estimate the $j^\text{th}$ moment. 
The relevant HG functions (grey, Eq.~\ref{eq:temdef}) are plotted for comparison: HG$_{k,0}(x,0)$ for the $2k^\text{th}$ moment; and $(\text{HG}_{k,0}(x,0) + \text{HG}_{k+1,0}(x,0) ) / \sqrt{2}$ for the $2k+1^\text{th}$ moment (see Ref.~\cite{Tsang2017}). 
For odd $j$ only the plus superpositions are shown for brevity; the minus superpositions are reflections about the origin.}
\label{fig:splicetem}
\end{figure*}

\begin{sidewaystable*}
  \begin{center}
    \caption{Overlaps of SPLICE projectors $\ket{\phi_j}$ with HG modes. The positions of the single-mode fiber are given by $f_j$, and the positions of the glass edges $\{ g_j  \}$ in units of $\sigma$. The positions for moments 8 and higher were solved numerically.  $\xi = \text{erf}^{-1} \left ( 1/2 \right )$. }
    \label{tab:splicetem}
    \begin{tabular}{c|c|c|c|c|c|c|c|c}
      & 
      $\ket{\phi_j}$ &
      $f_j / \sigma$ &
      $\{ g_j / \sigma \}$ &
      $\braket{\text{HG}_{10}}{\phi_j}$ &
      $\braket{\text{HG}_{20}}{\phi_j}$ &
      $\braket{\text{HG}_{30}}{\phi_j}$ &
      $\braket{\text{HG}_{40}}{\phi_j}$ &
      $\braket{\text{HG}_{50}}{\phi_j}$  \\
      \hline
      & & & & & & & \\
      $\mu_2$ & $\ket{\phi_2}$ & 0 & 0 & $\sqrt{\frac{2}{\pi}}$ & 0 & $- \frac{1}{\sqrt{3\pi}}$ & 0 & $\frac{1}{2} \sqrt{\frac{3}{5\pi}}$     
      \\
      & & & & & & & \\
      \multirow{2}{*}{$\mu_3$} &
      $\ket{\phi_3^+}$ &
      $\sqrt{2}$ &
      $1/\sqrt{2}$ &
      $\sqrt{\frac{2}{\pi e^{1/2}}}$ &
      $\sqrt{\frac{2}{\pi e^{1/2}}}$ &
      $ \frac{1}{2 \sqrt{3\pi e^{1/2}}}$ &
      $ -\frac{1}{\sqrt{6\pi e^{1/2}} }$ &
      $ - \frac{1}{8} \sqrt{\frac{3}{5\pi e^{1/2}}}$ 
      \\
      & $\ket{\phi_3^-}$ & $-\sqrt{2}$ & $-1/\sqrt{2}$ &  
      $\sqrt{\frac{2}{\pi e^{1/2}}}$  &
      $-\sqrt{\frac{2}{\pi e^{1/2}}}$ &
      $ \frac{1}{2 \sqrt{3\pi e^{1/2}}}$ &
      $ \frac{1}{\sqrt{6\pi e^{1/2}}}$ &
      $ - \frac{1}{8} \sqrt{\frac{3}{5\pi e^{1/2}}}$ 
      \\
      & & & & & & & \\
      $\mu_4$ & $\ket{\phi_4}$ &
      0 &
      $-\sqrt{2}\xi ,\sqrt{2}\xi$ &
      0 &
      $-2\sqrt{\frac{2}{\pi}}\xi e^{-\xi^2}$ &
      0 &
      $\sqrt{\frac{2}{3\pi}}\xi (3 - 2\xi^2) e^{-\xi^2}$ &
      0
      \\ 
      & & & & & & & \\
      \multirow{2}{*}{$\mu_5$} &
      $\ket{\phi_5^+}$ &
      $\frac{2}{\sqrt{3}}$ &
      $\frac{1}{\sqrt{3}} - \sqrt{2} \xi, \frac{1}{\sqrt{3}} + \sqrt{2} \xi$ &
      0 &
      $-2\sqrt{\frac{2}{\pi e^{1/3}}}\xi e^{-\xi^2}$ &
      $- 2\sqrt{\frac{2}{\pi e^{1/3}}}\xi e^{-\xi^2}$ &
      $\sqrt{\frac{2}{3\pi e^{1/3}}} \xi (1 - 2\xi^2) e^{-\xi^2}$ &
      $\sqrt{\frac{10}{\pi e^{1/3}}} \frac{\xi}{9} (7 - 6\xi^2) e^{-\xi^2}$ 
      \\
      & 
      $\ket{\phi_5^-}$ &
      $ -\frac{2}{\sqrt{3}}$ &
      $ -\frac{1}{\sqrt{3}} - \sqrt{2} \xi, -\frac{1}{\sqrt{3}} + \sqrt{2} \xi$ &
      0 &
      $ -2\sqrt{\frac{2}{\pi e^{1/3}}} \xi e^{-\xi^2} $ & $ 2\sqrt{\frac{2}{\pi e^{1/3}}} \xi e^{-\xi^2}$ & $ \sqrt{\frac{2}{3\pi e^{1/3}}} \xi (1 - 2\xi^2) e^{-\xi^2} $  & $-\sqrt{\frac{10}{\pi e^{1/3}}} \frac{\xi}{9} (7 - 6\xi^2) e^{-\xi^2}$
      \\
      & & & & & & & \\
      $\mu_6$ &
      $\ket{\phi_6}$ &
      0 &
      $-\sqrt{2 \ln 2}, 0, \sqrt{2 \ln 2}$ &
      0 &
      0 &
      $\frac{2 \ln 2}{\sqrt{3\pi}}$ &
      0 &
      $-\frac{2 \ln 2}{\sqrt{15\pi}}(3 - \ln 2)$
      \\
      & & & & & & & \\
      \multirow{2}{*}{$\mu_7$} &
      $\ket{\phi_7^+}$ &
      $1$ &
      $\frac{1}{2} - \sqrt{2 \ln 2}, \frac{1}{2}, \frac{1}{2} + \sqrt{2 \ln 2}$ &
      0 &
      0 &
      $\frac{2 \ln 2}{\sqrt{3\pi e^{1/4}}}$ &
      $\frac{2 \ln 2}{\sqrt{3\pi e^{1/4}}}$ &
      $-\frac{\ln 2}{2\sqrt{15\pi e^{1/4}}}(7 - 4 \ln 2)$
      \\
      & $\ket{\phi_7^-}$ &
      $-1$ &
      $-\frac{1}{2} - \sqrt{2 \ln 2}, -\frac{1}{2}, -\frac{1}{2} + \sqrt{2 \ln 2}$ &
      0 &
      0 &
      $\frac{2 \ln 2}{\sqrt{3\pi e^{1/4}} }$ &
      $-\frac{2 \ln 2}{\sqrt{3\pi e^{1/4}}}$ &
      $-\frac{\ln 2}{2\sqrt{15\pi e^{1/4}}}(7 - 4 \ln 2)$
      \\
      & & & & & & & \\
      $\mu_8$ &
      $\ket{\phi_8}$ &
      0 &
      -1.595, 0.509 , 0.509, 1.595 &
      0 &
      0 &
      0 &
      0.333 &
      0
      \\
      & & & & & & & \\
      \multirow{2}{*}{$\mu_9$} &
      $\ket{\phi_9^+}$ &
      0.894 &
      -1.148, -0.062, 0.956, 2.042 &
      0 &
      0 &
      0 &
      0.301 &
      0.301
      \\
      & $\ket{\phi_9^-}$ &
      -0.894 &
      -2.042, -0.956, 0.062, 1.148 &
      0 &
      0 &
      0 &
      0.301 &
      -0.301
      \\
      & & & & & & & \\
      $\mu_{10}$ &
      $\ket{\phi_{10}}$ &
      0 &
      -1.959, -0.933, 0, 0.933, 1.959 &
      0 &
      0 &
      0 &
      0 &
      0.243
    \end{tabular}
  \end{center}
\end{sidewaystable*}

\begin{table}
  \begin{center}
    \caption{Squared overlaps of SPLICE projectors $\ket{\phi_j}$ with $\ket{\psi_X}$, an emitter at position $X$ convolved with the point spread function. SPLICE measurements are centred on $\mu$, the mean of $F(X)$. Results are expanded to leading order in $\Delta = (X - \mu)/\sigma$.   $\xi = \text{erf}^{-1} \left ( 1/2 \right )$. }
    \label{tab:splicepsi}
    \begin{tabular}{c|c}
      $\ket{\phi_j}$ &  $|\braket{\phi_j}{\psi_X}|^2$  \\
      \hline
      & \\
      $\ket{\phi_2}$ & $\frac{1}{2\pi}\Delta^2 + \mathcal{O}(\Delta^4)$     
      \\
      & \\
      $\ket{\phi_3^\pm}$ & $\frac{1}{2 \sqrt{e} \pi} \left ( \Delta^2 \pm \frac{1}{2}\Delta^3 - \frac{1}{12}\Delta^4 \right ) + \mathcal{O}(\Delta^5)$     
      \\
      & \\
      $\ket{\phi_4}$ & $ \frac{1}{4}\xi^2 e^{-2\xi^2} \Delta^4 + \mathcal{O}(\Delta^6)$  
      \\ 
      & \\
      $\ket{\phi_5^\pm}$ & $ \frac{\xi^2 e^{-2\xi^2}}{4 e^{1/3} \pi} \left ( \Delta^4 \pm \frac{1}{\sqrt{3}}\Delta^5 - \frac{(5 - 2\xi^2)}{24} \Delta^6 \right ) +  \mathcal{O}(\Delta^7)$
      \\
      &\\
      $\ket{\phi_6}$ & $\frac{(\ln 2)^2}{288 \pi}\Delta^6 + \mathcal{O}(\Delta^8)$ 
      \\
      &\\
      $\ket{\phi_7^\pm}$ & $ \frac{(\ln 2)^2}{288 e^{1/4} \pi} \left ( \Delta^6 \pm \frac{1}{2}\Delta^7 - \frac{(11-2\ln2)}{40}\Delta^8 \right ) +  \mathcal{O}(\Delta^9)$
      \\
      & \\
      $\ket{\phi_8}$ & $(1.80\times 10^{-5}) \Delta^8 + \mathcal{O}(\Delta^{10})$
      \\
      & \\
      $\ket{\phi_9^\pm}$ & $(1.48\times 10^{-5}) \Delta^8 \pm (6.61\times 10^{-6}) \Delta^9 $
      \\
      & $- (3.99\times 10^{-6}) \Delta^{10} +  \mathcal{O}(\Delta^{11})$
      \\
      & \\
      $\ket{\phi_{10}}$ & $(4.60\times 10^{-7}) \frac{\Delta^{10}}{\sigma^{10}} + \mathcal{O}(\Delta^{12})$
    \end{tabular}
  \end{center}
\end{table}

\noindent So far we have focused on a particular scenario with two emitters. In this section we discuss how the SPLICE scheme can be adapted to estimate higher moments of a source distribution, enabling the characterization of more complex shapes. Here we take a more general distribution of point sources along the $x$-axis by considering the density matrix of an incoming photon as
\begin{eqnarray}
\rho &=& \int dX F(X) \ket{ \psi_X } \bra{\psi _X },
\\
\ket{\psi_X} &=& \iint dx dy \hspace{0.1cm} \psi(x - X,y ) \ket{x,y},
\nonumber
\end{eqnarray}
where we take the spatial extent of $F(X)$ to be less than the width of the point spread function ($\sigma$), firmly in the sub-Rayleigh regime.  
SPLICE measurement configurations are found using a similar method as in Section~\ref{sec:newsplice}: using $\pi$ phase-shifts and a single-mode fiber, each translatable along the $x$-axis, zero overlap with all lower HG modes can be obtained.   
We write a generalized SPLICE mode as  
\begin{eqnarray}
\label{eq:splicejjdef}
\ket{\phi_j} &=& \frac{1}{\sqrt{2 \pi \sigma^2}} \iint dx dy \hspace{0.1cm} \phi_j(x,y) \ket{x} \ket{y},
\\
\phi_j(x,y) &=& e^{-\frac{(x-f_j)^2 + y^2}{4\sigma^2}}  \prod_{i=1}^{n_j} \text{sgn} ( x - g_{j,i} ),
\nonumber
\end{eqnarray}
where $f_j$ is the corresponding position of the single-mode fiber, and the set $\{ g_j \}$ represent the positions of the right-most edges of $n_j$ $\pi$ phase-shifters, so that for every $x < g_j$ there is a phase shift of $\pi$.  
Figure~\ref{fig:splicetem} plots the SPLICE functions $\phi_j (x,0)$ for modes up to $j = 10$. 
Likewise, Table~\ref{tab:splicetem} details the fiber positions $f_j$, phase-shifter positions $\{ g_j \}$, and overlaps with HG modes up to $j=10$. 
Projective measurements onto the mode $\ket{\phi_j}$ succeed with probability
\begin{eqnarray}
p_j  &=&  \text{Tr}(\rho \ket{\phi_j}\bra{\phi_j}) = \int dX F(X) |\braket{\phi_j}{\psi_{X}}|^2.
\end{eqnarray}
By casting the SPLICE modes into the HG basis, we can use the same treatment as in Ref.~\cite{Tsang2017} to show how the $j^\text{th}$ statistical moment of $F(X)$ is extracted in the sub-Rayleigh regime. Here we will consider when the SPLICE measurements are centred about $\mu_1$, the mean of $F(X)$, which is to say that the specified positions of the fiber and phase-shifters in Table~\ref{tab:splicetem} should be made relative to $\mu_1$. As in the previous section, intensity measurements can be used to efficiently estimate the mean. 

Table~\ref{tab:splicepsi} shows the probability of success for a projective measurement onto the SPLICE mode $\ket{\phi_j}$. The results show leading order terms in a series expansion of $\Delta = (X-\mu_1)/\sigma$, the distance from the point source at $X$ to the mean of $F(X)$. It can be seen from Table~\ref{tab:splicepsi} that $| \braket{\phi_j}{\psi_X}|^2 \propto \Delta^j$  (for odd $j$ this is true once the difference $| \braket{\phi^+_j}{\psi_X}|^2 - | \braket{\phi^-_j}{\psi_X}|^2$ is taken). This directly implies that the integral in $p_j$ gives the expectation value of $(X-\mu_1)^j$, which is the $j^\text{th}$ moment of $F(X)$ centred on the mean. This is true for any choice of $F(X)$ in the sub-diffraction regime. The distribution $F(X)$ is uniquely determined by its moments. Measuring up to the 10th moment one can, in principle, completely characterize distributions with 10 or fewer degrees of freedom, e.g., 5 sources with unequal intensities.
Though the SPLICE scheme is capable of estimating arbitrary moments of $F(X)$, we note here that higher moments require substantial precision in the placement of the phase-shifters. For example, the $\ket{\phi_{10}}$ projection shows competitive contributions from lower orders of $\Delta/ \sigma$ if the values of $\{ g_{10} \}$ differ from ideal by more than $10^{-3} \sigma$. One could imagine achieving this level of precision for space telescopes which have point spread functions on the metre-scale.

\section{Conclusions}
\label{sec:discussion}
\noindent The SPLICE measurement scheme, initially envisioned to demonstrate super-resolution in a highly idealized scenario, can also be advantageous in a more realistic system, such as where the point sources have unequal intensities. 
Surprisingly, the SPLICE measurement, without any modification, outperforms IPC for resolving lower separations at moderately disparate intensities. Eventually, the skew of the source distribution becomes so substantial that the SPLICE measurement retains a large bias.
But in this case a simple adaptation to SPLICE regains an advantage in resolving small separations over IPC. 
We found, by comparing Cram{\' e}r-Rao bounds, that the advantage over IPC persists to very imbalanced intensities (at least $q = 10^{-4}$), lending some credence to the use of novel sub-Rayleigh imaging techniques for exoplanet detection (Earth is $10^{10}$ times fainter than the Sun at visible wavelengths~\cite{Seager2014}).
The adapted SPLICE scheme can be realized without increasing the complexity of the scheme: the only requirements remain a $\pi$ phase-shift about an edge and a single-mode fiber, now translatable along the transverse axis. 
Thus, we imagine that an experimental demonstration of sub-Rayleigh imaging for emitters of very different brightness can be performed in much the same style as Ref.~\cite{Tham2017}. 
We have focused on a source distribution in one dimension, however, it is a straightforward extension to also include measurement along the $y$-axis to investigate two-dimensional distributions. 
Finally, we have proposed extensions to the SPLICE scheme beyond measurements of the variance and skew, up to the $10^\text{th}$ statistical moment.
These measurements add some complexity to the scheme, requiring multiple edges between regions of $\pi$ and zero phase-shifts. 
We imagine that this could be performed with a spatial light modulator followed by an array of single-mode fibres. 
The ability to measure higher statistical moments of a source distribution enables one to reconstruct spatial structure. 


\begin{acknowledgments}
The authors would like to thank Mankei Tsang, John Donohue, Noah Lupu-Gladstein and Arthur Ou Teen Pang for fruitful discussions. This work was supported by Natural Sciences and Engineering Research Council (NSERC) of Canada and from the Canadian Institute for Advanced Research (CIFAR).
\end{acknowledgments}

\bibliography{imaging}
\appendix

\section{Monte Carlo simulations}
\label{sec:MC}

Monte Carlo simulations of SPLICE, adapted SPLICE, and IPC were performed using \emph{Wolfram Mathematica 11}. For all simulations the PSF width is set as $\sigma = 1$. 
First, we describe the IPC simulations. 
The inputs are the actual values of $\delta$, $q$ and $c$ as well as the number of incident photons $N$. 
We then draw $N$ position values $x_i$ from the intensity distribution
\begin{equation}
\mathcal{I}(x) = q \frac{1}{\sqrt{2 \pi \sigma^2}} e^{- \frac{(x - (c + \delta/2))^2}{2 \sigma^2}} + (1-q) \frac{1}{\sqrt{2 \pi \sigma^2}} e^{- \frac{(x - (c - \delta/2))^2}{2 \sigma^2}}.
\end{equation}
Note that we take the centre of the distribution to be $c=0$ throughout the simulations, since the SPLICE measurement is independent of $c$ when aligned about the mean of $\mathcal{I}(x)$.
Next, we estimate the mean, variance (subtracting $\sigma^2$), and skew of the resulting set of position values $\{ x \}$, 
\begin{eqnarray}
\bar x &=& \frac{1}{N} \sum_i x_i, 
\\
\hat v &=& \frac{1}{N-1} \sum_i (x_i - \bar x)^2  - \sigma^2,
\\
\hat s &=& \frac{N}{(N-1)(N-2)} \sum_i (x_i - \bar x)^3.
\end{eqnarray}
The estimates $\hat \delta$ and $\hat q$ are then calculated using Eqs.~\ref{eq:deltaest} and \ref{eq:qest}.
Importantly, we rejected Monte Carlo trials where $\hat v< 0$ since it led to unphysical estimates of $\delta$. This is a substantial cause of the bias in the IPC measurement as low $N$.
We note that other methods, such as regression, for estimating $\delta$ using IPC were explored but did not perform better than inverting the variance and skew as above. 

Next, we outline how the Monte Carlo simulations for SPLICE, and adapted SPLICE, were performed. For the variance measurement the probability of a photon detection, as a function of $\delta, q, c$ and the position of the fiber and phase-shifter edge ($f$ and $g$, respectively), is 
\smaller
\begin{equation}
P =  q \left[ e^{-\frac{f-(c+\delta/2))^2}{8\sigma^2}} \text{erf} \left ( \frac{f-2g + c + \delta/2}{2\sqrt{2} \sigma} \right ) \right]^2 
+ (1-q) \left[ e^{-\frac{f-(c-\delta/2))^2}{8\sigma^2}} \text{erf} \left ( \frac{f-2g + c - \delta/2}{2\sqrt{2} \sigma} \right ) \right]^2.
\label{eq:MCprob}
\end{equation}
\normalsize
Notice that when aligned at the mean, $f=g= \mu_1 = c - (1-2q)\delta/2$, that $c$ drops out entirely. 
To estimate the variance, given an average of $N$ incident photons, we draw a number $k$ from a Poisson distribution with $\lambda = N \times P$.
The variance is then estimated as $\hat v = 2 \pi \sigma^2 \left( \frac{k + 1/2}{N + 1} \right )$.
The factors of $1/2$ and $1$ are added to the numerator and denominator so that the variance is never estimated to be 0.
For the first-generation SPLICE scheme the separation is estimated as $\sqrt{4 \hat v}$.
When the mean of the distribution is not known a priori, $M$ photons are detected with IPC to estimate it, and so the values of $f$ and $g$ are set equal to $\bar x$.

For the adapted SPLICE scheme we add two measurements to estimate the skew. 
The probabilities of photon detection, $p_3^+$ and $p_3^-$, remain the same as Eq.~\ref{eq:MCprob}, where the fiber and phase-shifter edge are moved to $f = \bar x + \sqrt{2} \sigma$, $g = \bar x + \sigma/\sqrt{2}$ for the first measurement, and $f =  \bar x - \sqrt{2} \sigma$, $g = \bar x - \sigma/\sqrt{2}$ for the second. 
Incident photons ($N$), are evenly divided between these two measurements, and we draw two numbers, $k_+$ and $k_-$, from Poisson distributions with means $\lambda_\pm = N/2 \times p_3^\pm$.
The skew estimate is calculated as $\hat s = \sqrt{2 e} \pi \sigma^3 \left ( \frac{k_+ - k_-}{N/2} \right)$.
We typically allocate incident photons evenly between the measurements such that $N/3$ are used to estimate the mean with IPC, $N/3$ are used to estimate the variance, and $N/6$ are used in each of the skew measurements.

\section{Expectation value of SPLICE estimator in certain limits}
\label{sec:numerical}

The number of photon detections ($k$) using the SPLICE measurement ($\ket{\phi_2}$), for a fixed number of input photons ($N$), is drawn from a binomial distribution with probability $p_2 = \bra{\phi_2}\rho\ket{\phi_2}$. 
However, since we take $N$ to be a Poissonian average number of incident photons, $k$ is then drawn from a Poisson distribution with mean $\lambda = N \times p_2 \approx \frac{N \delta^2}{8 \pi \sigma^2}$ in the simplest case.
Since the probability of detection depends on $\delta^2$ to leading order, the estimator for the separation ($\hat \delta_\text{SPL}$) depends on $\sqrt{k}$, specifically $\hat \delta_\text{SPL} (k) = \sqrt{8 \pi \sigma^2 k/ N}$.
We are then interested in the expectation value $E \left [\sqrt{k} \right ] = \sum_k \sqrt{k} \frac{e^{-\lambda} \lambda^k}{k!}$ in the low, and high, limits of $N$.
For $\lambda = N \times p_2 \ll 1$, the first few terms in the sum give a good approximation to the expectation value, and we have
\begin{eqnarray}
E \left [\sqrt{k} \right ] & = & \left (1 - \lambda + \frac{\lambda^2}{2} - ...  \right )  
\left ( \lambda + \frac{\lambda^2}{\sqrt{2}} + ...  \right )
\nonumber \\ 
& \approx & 
\lambda - \left ( 1-\frac{1}{\sqrt{2}} \right ) \lambda^2.
\end{eqnarray}
The bias is $\approx \sqrt{\lambda}(\sqrt{\lambda} - 1)$, and, since $ E \left [ k \right ] = \lambda$, the variance is $V \left [ \sqrt{k} \right ] \approx \lambda$.  Notably, this implies that for low $N$, the variance of the SPLICE estimator is constant: $V \left [ \hat \delta_\text{SPL} \right ] = \delta^2$. This behaviour continues until $\lambda \sim 1$.  We consider $N$ to be large when $\lambda = N \times p_2 \gg 1$. In this case we can expand the estimator around its mean value,
\begin{eqnarray}
\sqrt{k} & = & 
\sqrt{\lambda} 
+ \frac{1}{2 \sqrt{\lambda}} (k - \lambda)
- \frac{1}{8 \lambda^{3/2}} (k - \lambda)^2
+ ... 
\nonumber \\
E \left [ \sqrt{k} \right ] & \approx & 
\sqrt{\lambda} 
- \frac{1}{8 \sqrt{\lambda}}.
\end{eqnarray}
The bias of the SPLICE estimator in this case is $\frac{\sqrt{\pi} \sigma}{N \delta}$, and the variance is $V \left [ \hat \delta_\text{SPL} \right ] = \frac{2 \pi \sigma^2}{N}$.

When the mean of the position is not known a priori, so that the SPLICE measurement may be misaligned, the probability of a photon detection picks up an extra term which must be taken into account when calculating the expectation value of the SPLICE estimator. Similarly, when the equal-intensity assumption is relaxed the probability of detection is modified, becoming 
\begin{equation}
\lambda(\Delta\mu) \approx N\frac{q(1-q)\delta^2}{2\pi \sigma^2} + N \Delta \mu \frac{2q(1-q)(1-2q)\delta^3}{3\pi \sigma^4} + N \frac{\Delta \mu^2}{2\pi \sigma^2} \left (1 - 2q(1-q)\frac{\delta^2}{\sigma^2} \right ).
\end{equation} 
The misalignment $\Delta \mu$ is a random normal variable with mean 0, and variance $\sigma^2 / M$, where $M$ is the number of photons used for estimating the mean with a round of IPC. To calculate the expectation value $E \left [ \hat \delta_\text{SPL} \right ]$ we must integrate over the probability density of $\Delta \mu$ as well as take the Poissonian sum. In the main text, we only do this for the large $N$ case,
\begin{eqnarray}
E \left [ \hat \delta_\text{SPL} \right ]=  \sqrt{\frac{8\pi \sigma^2}{N}} \int_{-\infty}^{\infty} \left ( \sqrt{\lambda(\Delta \mu)} - \frac{1}{8 \sqrt{\lambda(\Delta \mu)}} \right ) \frac{e^{-\frac{\Delta \mu^2}{2 \sigma^2/M}}}{\sqrt{2 \pi \sigma^2/M}}  d(\Delta \mu),
\end{eqnarray}
which we integrate numerically.
For large $N$, and subsequently large $M$, $E \left [ \hat \delta_\text{SPL} \right ] \propto \delta$ and $E \left [ \hat \delta_\text{SPL}^2 \right ] \propto \delta^2$.

\section{Cram{\' e}r-Rao bounds for SPLICE and IPC}
The Cram{\' e}r-Rao bound (CRB) for a parameter estimate is calculated by inverting the Fisher information matrix.  
The Fisher information matrix $\mathcal{F}_a$, given $P$, the probability of getting a certain result $a$, and the parameters $\delta$, $q$, and $c$ is
\begin{equation}
\mathcal{F}_a = \frac{1}{P}
\begin{pmatrix}
P_{\delta \delta} & P_{\delta q} & P_{\delta c}
\\
P_{q \delta} & P_{q q} & P_{q c}
\\
P_{c \delta} & P_{c q} & P_{c c}
\end{pmatrix}.
\end{equation}
For IPC, $P = \mathcal{I}(x)$, the intensity distribution. The full Fisher information matrix $\mathcal{F}$ is then integrated over all measurement outcomes $\mathcal{F} = \int_{-\infty}^{\infty} \mathcal{F}_x dx$. 
The resulting matrix is then inverted with the first entry corresponding to the CRB for estimating $\delta$. 
We perform this integration and matrix inversion numerically in the main text. 

For SPLICE, $P = \frac{1}{2} p_2$ for the variance measurements, and $P = \frac{1}{4} p_3^+,  \frac{1}{4} p_3^-$ for the skew measurement.
The factors of $1/2$ and $1/4$ represent the fact that photons are allocated evenly between the variance and skew measurements.
We assume that the mean is known ($\Delta \mu = 0$) and so do not include the round of IPC for mean estimation here.
This is addressed by scaling the photon number $N$ by a factor of $2/3$ to account for $1/3$ of incoming light being allocated to IPC for the mean estimate. 
The full Fisher information matrix is then $\mathcal{F} = \frac{1}{2} \mathcal{F}_{p_2} + \frac{1}{2} \mathcal{F}_{1-p_2} + \frac{1}{4} \mathcal{F}_{p_3^+} + \frac{1}{4} \mathcal{F}_{1-p_3^+} + \frac{1}{4} \mathcal{F}_{p_3^-} + \frac{1}{4} \mathcal{F}_{1-p_3^-}$.
Once again, we evaluate this numerically for the different values of $\delta$ and $q$, and invert the resulting matrix to find the CRB for estimating $\delta$. 

\end{document}